\documentclass[aps,prl,twocolumn,superscriptaddress]{revtex4}
\usepackage{graphicx}% Include figure files
\usepackage{dcolumn}% Align table columns on decimal point
\usepackage{bm}% bold math
\usepackage{url}
\begin{document}

% Title and Abstract
%---------------------------------------------------------------------
\title{The regulatory network of {\it E. coli} metabolism as a boolean
dynamical system exhibits both homeostasis and flexibility of
response}

\author{Areejit Samal}
\affiliation{Department of Physics and Astrophysics, University of
Delhi, Delhi 110007, India}
\author{Sanjay Jain}
\email[Corresponding author:]{jain@physics.du.ac.in}
\affiliation{Department of Physics and Astrophysics, University of
Delhi, Delhi 110007, India} \affiliation{Jawaharlal Nehru Centre for
Advanced Scientific Research, Bangalore 560064} \affiliation{Santa
Fe Institute, 1399 Hyde Park Road, Santa Fe, NM 87501, USA}
%---------------------------------------------------------------------

\begin{abstract}
Elucidating the architecture and dynamics of large scale genetic
regulatory networks of cells is an important goal in systems
biology. We study the system level dynamical properties of the
genetic network of {\it Escherichia coli} that regulates its
metabolism, and show how its design leads to biologically useful
cellular properties. Our study uses the database (Covert {\it et
al.}, Nature 2004) containing 583 genes and 96 external metabolites
which describes not only the network connections but also the
boolean rule at each gene node that controls the switching on or off
of the gene as a function of its inputs. We have studied how the
attractors of the boolean dynamical system constructed from this
database depend on the initial condition of the genes and on various
environmental conditions corresponding to buffered minimal media. We
find that the system exhibits homeostasis in that its attractors,
that turn out to be fixed points or low period cycles, are highly
insensitive to initial conditions or perturbations of gene
configurations for any given fixed environment. At the same time the
attractors show a wide variation when external media are varied
implying that the system mounts a highly flexible response to
changed environmental conditions. The regulatory dynamics acts to
enhance the cellular growth rate under changed media. Our study
shows that the reconstructed genetic network regulating metabolism
in {\it E. coli} is hierarchical, modular, and largely acyclic, with
environmental variables controlling the root of the hierarchy. This
architecture makes the cell highly robust to perturbations of gene
configurations as well as highly responsive to environmental
changes. The twin properties of homeostasis and response flexibility
are achieved by this dynamical system even though it is not close to
the edge of chaos.
\end{abstract}
\maketitle

%-------------------------------------------------------------------

\section*{Introduction}

\noindent Large scale biological networks and their associated
dynamical systems have a crucial role to play in unravelling the
systemic properties of cells. Structural studies of large scale
metabolic, protein interaction and genetic regulatory networks have
uncovered some unexpected patterns leading to interesting hypotheses
and questions (for reviews see \cite{BO2004,Alon2003,BLGS2004}). For
a deeper understanding of system level phenomena, it now seems that
we need to explore the relationship between network structure and
the dynamics of genes, proteins and other biomolecules. In this
paper we study the {\it Escherichia coli} regulatory network and
show that the dynamics leads to biologically important properties
such as cellular homeostasis and flexibility of response to varied
environments. Our study reveals that some very simple features of
the genetic regulatory network are responsible for these properties.
These design features may be universal across prokaryotes and
possibly have vestiges in higher organisms as well.

Large scale mathematical models for dynamical phenomena are
difficult to construct due to paucity of data and are difficult to
profitably analyze due to their complexity. In this context flux
balance analysis (FBA) has proved to be a useful computational
technique to explore steady state flows in large scale metabolic
networks \cite{VP1994,EP2000,EIP2001,SVC2002}. A conceptual
framework to study dynamics of large scale genetic regulatory
networks as boolean systems was introduced by Kauffman
\cite{K1969a,K1969b,OriginsOfOrder}. In this paper we use this
approach to study the large scale transcriptional regulatory network
(TRN) of an organism in which both the network and the boolean
functions have been constructed from real data. Our study is based
on the database iMC1010v1 \cite{CKRHP2004} which describes the
regulatory network controlling metabolism in {\it E. coli}.

The boolean approach provides a coarse-grained model of the dynamics
of TRNs, in which each gene's configuration has only two allowed
values (corresponding to the gene being off or on), each gene's
update is given by a boolean function of all its inputs, time is
discrete and (in our work) all genes are updated synchronously. A
differential equation based simulation of large scale TRNs is not
feasible at the moment due to lack of kinetic data, and the large
number of unknown parameters would also render the results of such a
simulation difficult to interpret \cite{Bornholdt2005}. On the other
hand boolean simulations of smaller biological systems have provided
useful insights \cite{ST2001,AO2003,LLLOT2004,EPA2004,LAA2006}. The
boolean approach can provide useful information about some
qualitative features of the dynamics, e.g., the nature of the
attractors of the system, and through that, insights about what
might happen in a more detailed simulation and the system itself.

%----------------------------------------------------------------------------------------------

\begin{figure*}
\centering
\includegraphics[width=16cm]{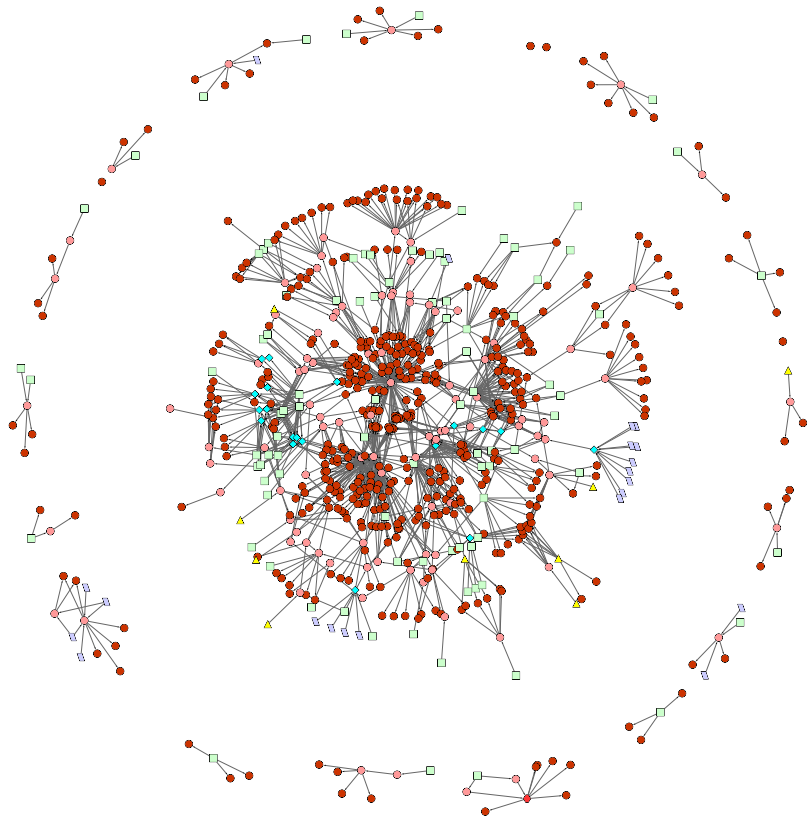}
\caption{Map of the transcriptional regulatory network controlling
metabolism in {\it E. coli}. In this figure, there are genes coding
for the TFs (pink circles), genes coding for enzymes (brown
circles), external metabolites (green squares), certain internal
fluxes (purple parallelograms), stimuli (yellow triangles) and other
conditions (blue diamonds). See text for details. }
\end{figure*}

\begin{figure}
\centering
\includegraphics[width=7cm]{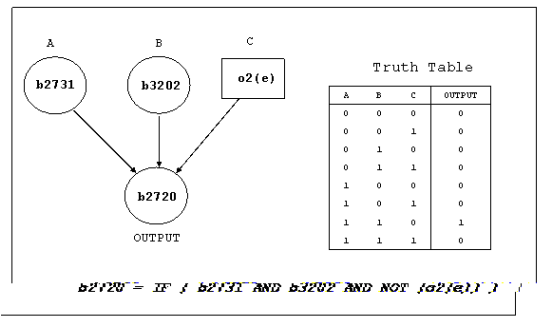}
\caption{Example of a boolean function $G_i$ representing the
regulatory logic at the promoter region of gene b2720 that
determines its expression. The gene b2720 is on if and only if both
the transcription factors coded by genes b2731 and b3202 are present
and oxygen is absent in the environment. For all other cases, the
gene b2720 is off.}
\end{figure}

\subsection*{The genetic network regulating {\it E. coli} metabolism
as a boolean dynamical system}

The database iMC1010v1 contains 583 genes. These are collectively
regulated by a set of 103 transcription factors (TFs) which are gene
products of 104 of the genes in the set, 96 external metabolites, 19
other conditions, 21 internal fluxes of metabolic reactions and 9
stimuli. The directed graph of this network is shown in Fig. 1,
where a directed link from one node to another denotes a regulatory
interaction.

The database also provides the boolean input-output map at each
node, e.g., the configuration of each gene (on or off), as a
function of the on-off states of all its inputs. Using this
information we construct the following discrete dynamical system
describing {\it E. coli's} TRN (for details, see Methods section):
\begin{equation}
g_i(t+1) = G_i({\bf g}(t), {\bf m}); \quad \quad i = 1, 2, \ldots,
583.
\end{equation}
Here $g_i(t)$ is the configuration of gene $i$ at time $t$. Time is
measured in discrete units: $t = 0,1,2,\ldots$. $g_i(t)$ = 1 (0)
means that at time $t$ gene $i$ is on (off). The vector ${\bf g}(t)$
collectively denotes the configurations of all the genes at time
$t$; its $i^{th}$ component is $g_i(t)$. The vector ${\bf m}$
denotes the configuration of external metabolites; its $i^{th}$
component $m_i$ = 1 if metabolite $i$ ($i=1,2,\ldots,96$) is present
in the external environment for uptake into the cell, and $m_i = 0$
if it is absent. The above equation expresses the fact that the
on-off state of a gene at any time instant is controlled by the
state of the genes at the previous time instant as well as the state
of the external environment. The interaction of genes is mediated by
transcription factors. Thus a single time unit corresponds to the
average time between the initiation of transcription of a gene
coding for a transcription factor and the initiation of
transcription of a gene regulated by that transcription factor.

In principle ${\bf m}$ can also change with time as the cell uses up
food molecules in its external environment for its metabolism and
excretes other molecules \cite{CSP2001,BHRP2005}. However, in the
present work we consider only buffered media which are characterized
by $m_i$ that are constant in time. ${\bf m}$ thus defines a
constant external environment of the cell. We have considered two
classes of buffered media, (a) a set of 93 minimal media (62 aerobic
and 31 anaerobic) each capable of supporting the growth of the cell
as determined by FBA (see Table S1 for a list), and (b) a much
larger library of 109732 minimal media constructed using the method
described by Barrett et al \cite{BHRP2005}.

The functions $G_i$ contain all information about the internal
wiring of the network (who influences whom) as well as the logic of
each gene's regulation (given the configuration of all of gene $i$'s
inputs at time $t$, whether gene $i$ will be on or off at $t+1$).
Each function $G_i$ typically depends only upon those components of
${\bf g}$ and ${\bf m}$ that directly affect the expression of gene
$i$ (see Fig. 2 for an example). We have considered the dynamical
system (1) with two slightly different forms of the functions $G_i$,
called 1A and 1B, arising from two different treatments of
intermediate variables (the internal fluxes of certain metabolic
reactions) that appear in the database iMC1010v1. In the first
approach (1A) for simplicity we have treated only the genes and
their products as dynamical variables, keeping these internal fluxes
fixed. The second approach (1B) includes the effect of some other
internal variables such as concentrations of internal metabolites
(as reflected through these fluxes) also being dynamical. The latter
effectively introduce additional interactions among the genes.

The conceptual framework for studying TRNs as boolean dynamical
systems of the type $g_i(t+1) = G_i({\bf g}(t))$ was set up by
Kauffman \cite{K1969a,K1969b} almost four decades back and
subsequently has been studied extensively, resulting in several
important insights (see, e.g.,
\cite{Thomas1973,OriginsOfOrder,HSWK2002,SDKZ2002,KPST2003,KPST2004}).
In particular Kauffman found that such systems with a large number
of components possess an ordered regime in which the attractors have
short periods and large basins. In this regime these systems have
the property of homeostasis or robustness to perturbations of the
genetic configuration. In the absence of detailed molecular data on
the real genetic networks, this approach was used for ensembles of
biologically motivated random boolean networks, and, more recently,
real networks with the functions $G_i$ chosen randomly from a
suitable ensemble of boolean functions\cite{KPST2003,KPST2004}.

References \cite{ST2001,AO2003,LLLOT2004,EPA2004,LAA2006} have
applied the boolean approach to specific biological gene regulatory
networks where detailed genetic data is available. These networks
are smaller than the ones mentioned above, and have up to 40
distinct genes, proteins and other molecules
\cite{ST2001,AO2003,LLLOT2004,EPA2004,LAA2006}. In reference
\cite{AO2003}, where a boolean network of 180 nodes is considered,
the network contains 15 distinct genes and proteins (with 12 nodes
for each of them corresponding to 12 distinct cells). These models,
apart from reproducing several observed phenomena of these systems,
have also found that the networks possess the property of
homeostasis, as well as robustness to genetic mutations.

The present study is inspired by the work of Kauffman and extends
the above development in two important ways. One, it studies the
empirically derived network of a real organism, but one that much
larger than the biological systems mentioned above. The present
network \cite{CKRHP2004} has 583 genes and 96 external metabolites
accounting for close to half of all genes currently believed to be
involved in metabolism in {\it E. coli}. Being more than an order of
magnitude larger (in terms of the number of genes involved) than
other real genetic networks considered as boolean systems, this
allows us a qualitatively different systemic view of the
organization of the genetic network of an organism. We not only find
homeostasis in this large system, but also identify the design
feature of the network responsible for this property. Two, we are
able to study the effect of the external environment on the TRN
dynamics through the vector ${\bf m}$ in Eq. (1) in a much more
systematic and extensive way than before. This sheds light on a
different property of the network, namely its flexibility of
response to a diversity of environments.

%----------------------------------------------------------------------------------------------

\section*{Results}

%----------------------------------------------------------------------------------------------

\subsection*{Homeostasis: The final state is essentially the same after any
perturbation of the genes}

We simulated the dynamical system 1A for each of the 93 ${\bf m}$
vectors corresponding to the 93 minimal media mentioned above,
starting from a set of 10000 randomly chosen initial conditions for
the $g_i$. For each ${\bf m}$ and each initial condition of the
genes, the system reached a fixed point attractor in a maximum of 4
time steps. Furthermore, for each ${\bf m}$ the fixed point was
independent of the chosen initial condition of the genes. This is
shown in Fig. 3 for glucose aerobic medium for four initial
conditions. We also considered the library of 109732 minimal media
for a single randomly chosen initial condition each. A fixed point
attractor was found in each case. There are in principle $2^{583}$
possible initial conditions. We present later the analytic argument
as to why a unique final configuration independent of initial
condition is inevitable for each fixed ${\bf m}$, given the
architecture of the TRN. This property means that as long as the
external environment remains fixed, the TRN regulating {\it E. coli}
metabolism will revert to a unique configuration of its genes after
any perturbation of the latter.

The dynamical system 1B, which includes some additional links
between the genes compared to 1A, was also studied for the 93
minimal media with 1000 randomly chosen initial conditions each. In
this case for $89$ of the $93$ media, we found $36$ distinct
attractors ($8$ fixed point attractors and $28$ two-cycles). For the
remaining $4$ minimal media, there were $10$ distinct attractors
($4$ fixed point attractors and $6$ two-cycles). Again the attractor
was reached in a maximum of 4 time steps. For each of the cycles, we
found that most of the genes (562 to 567 out of 583) were in fact
locked in a fixed configuration, and only 16 to 21 genes oscillated
back and forth between zero and one with period two. Kauffman has
characterized random boolean networks as having two regimes, an
ordered regime wherein the attractors have a large `frozen core' of
genes locked in a fixed configuration together with a few `twinkling
islands' of genes that switch on and off, and a chaotic regime
wherein the number of `frozen' genes is much less than those of the
`twinkling' ones \cite{OriginsOfOrder}. Our findings above are
consistent with Kauffman's hypothesis that real genetic regulatory
networks are in the ordered regime.

Furthermore for any given medium we found that each of the frozen
genes had the same configuration across all the attractors (36 or
10). This means that for any given medium, most genes (562 or more
out of 583) end up in the same fixed configuration independent of
the initial conditions of the genes. It can analytically be checked
that there are no other attractors of this system, using its
structural properties.

Collectively, our results of both dynamical systems imply that the
{\it E. coli} TRN exhibits a high degree of homeostasis, in that it
is highly insensitive to initial conditions and for any given medium
all genetic perturbations die out quickly, restoring an overwhelming
majority of genes to a configuration that is independent of the
perturbation.

%----------------------------------------------------------------------------------------------

\begin{figure}
\centering
\includegraphics[width=7cm]{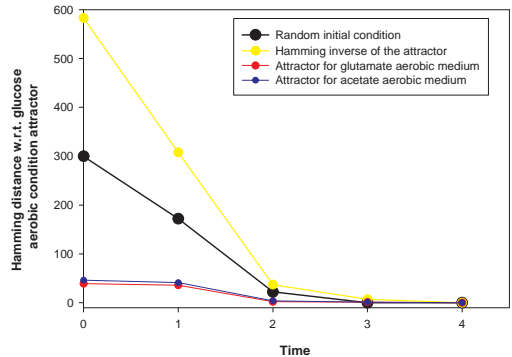}
\caption{Dynamical behaviour of the {\it E. coli} TRN for a fixed
environment, glucose aerobic minimal media. For all initial
conditions the system is attracted to a fixed point whose
configuration depends upon the medium. The plots depict, as a
function of time, the hamming distance of the configuration from the
fixed point attractor corresponding to the medium. 4 different
initial conditions are shown. One is a randomly chosen initial
condition. Another is the `hamming inverse' of the attractor (in
which the configuration of every gene is reversed with respect to
the attractor). Two other initial conditions are the attractor
configurations of other minimal media.}
\end{figure}

\begin{figure}
\centering
\includegraphics[width=7cm]{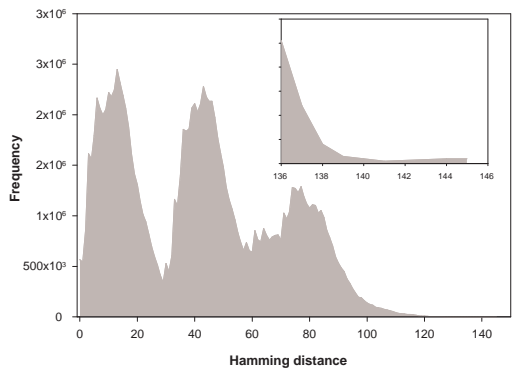}
\caption{The {\it E. coli} TRN is flexible in response to changing
environmental conditions encountered. Changing the environmental
condition can lead to a wide range of hamming distances among the
attractors. In the figure, the distribution of pair-wise hamming
distances between attractors for 15,427 different environmental
conditions is shown. Inset: Enlargement of the graph for large
hamming distances. The largest hamming distance obtained between
attractors for two different environmental conditions is 145. }
\end{figure}

\begin{figure}
\centering
\includegraphics[width=7cm]{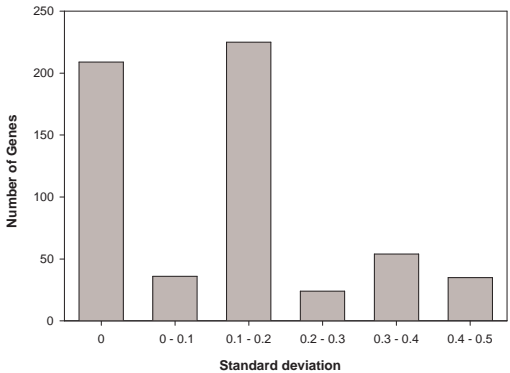}
\caption{The histogram of standard deviation of a gene's
configurations across 15427 attractors for different environmental
conditions. The left-most bar corresponds to 209 genes whose
configuration remains unchanged.}
\end{figure}

\subsection*{Flexibility: The system has a wide range of response
to changes in environmental conditions}

While homeostasis is a useful property in any given environmental
condition, the organism also needs to respond flexibly to changes in
the environment. We investigated flexibility of the TRN to
environmental changes in two ways. First, we determined the hamming
distance between attractor states of the system 1A corresponding to
pairs of minimal media. For the set of 93 minimal media, we found
the largest hamming distance between two attractor states
corresponding to two different minimal media to be $114$. We also
determined the attractors of the dynamical system 1A for the larger
library of 109732 minimal media (all attractors are fixed points
whose basin of attraction is the entire configuration space). We ran
constrained FBA for each of these attractors to determine which of
them supports a nonzero growth rate (see Methods section for
details). This yielded a subset of 15427 minimal media. We computed
the pairwise hamming distances among this set of 15427 attractors
also. The largest of these distances was found to be 145. The
distribution of these hamming distances is trimodal similar to that
found by \cite{BHRP2005}, and is shown in Fig. 4. Thus, although the
attractor for a fixed environmental condition is unique, the
attractors for two different environmental conditions can be quite
far apart. Therefore, while the system is insensitive to
fluctuations in gene configurations in a fixed external environment,
it can move to quite a different attractor when it encounters a
change in environment. Thus the system shows flexibility of response
to changing environmental conditions.

Second, we found that across these 15427 conditions the genes that
had a configuration that differed between any pair of attractors
were drawn from a set of 374 out of the 583 genes. The remaining 209
genes had the same configuration (75 off and 134 on) in all the
15427 attractors. The variability of a gene's configuration across
different environmental conditions can be characterized by the
standard deviation of its value (zero or one) across this set. We
found this standard deviation to range from zero to close to its
maximum possible value 0.5, with the mean of the 374 standard
deviations mentioned above being 0.20. The histogram of standard
deviation values is shown in Fig. 5. These observations quantify the
considerable variety in a gene's variability across environmental
conditions.

%----------------------------------------------------------------------------------------------
\begin{figure}
\centering
\includegraphics[width=7cm]{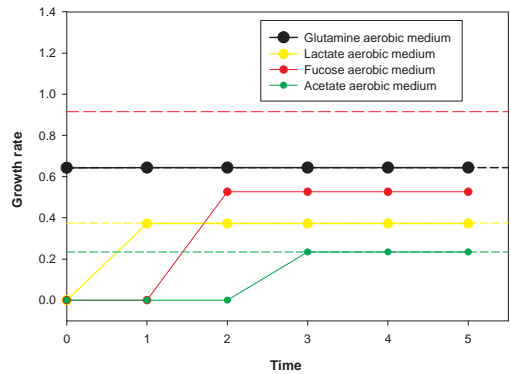}
\caption{Metabolic efficiency due to regulation. The figure shows
the adaptation of the {\it E. coli} TRN towards higher growth rate
in response to change of medium. Growth rate obtained using
constrained FBA is plotted for 4 trajectories of the TRN
corresponding to aerobic minimal media with glutamine, lactate,
fucose or acetate as the carbon source. The initial condition of the
TRN in each case is the state the system would have been in for the
glutamate aerobic medium. Dotted lines show the pure FBA growth rate
in the 4 minimal media. The growth rate increases in three and
remains constant in one of these trajectories.}
\end{figure}

\begin{figure}
\centering
\includegraphics[width=7cm]{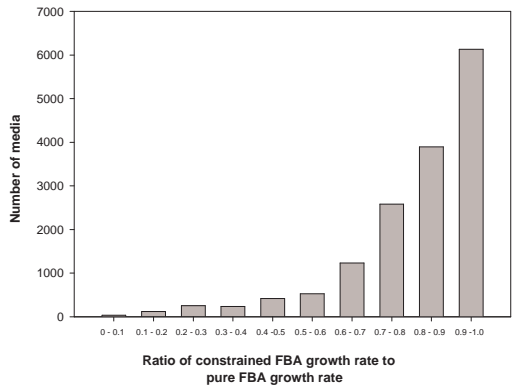}
\caption{Histogram of the ratio of constrained FBA growth rate in
the attractor of each of 15427 minimal media discussed in text to
the pure FBA growth rate in that medium. This is peaked in the bin
with the largest ratio ($\geq 0.9$).}
\end{figure}

\subsection*{Adaptability: The genetic network's response to
changed media increases metabolic efficiency}

To further investigate flexibility, we tracked how the metabolic
response of the cell, as measured by its growth rate computed using
FBA, changes when its environment changes. A reaction in the
metabolic network can be assumed to be off if none of the enzymes
catalyzing it are being produced, or, equivalently, in our dynamical
system, if the genes coding for those enzymes are in the off state.
For any configuration of the metabolic genes, FBA can thus be used
to compute the growth rate of the cell by turning off all reactions
whose corresponding genes are in the off state in that
configuration, thereby capturing the effect of gene regulation on
metabolic function (see Methods section). We computed this
`constrained FBA' growth rate for each of the attractors of the TRN
dynamical system 1A for the 93 minimal media. 81 of them, listed in
Table S2 in Additional File 1, gave a nonzero growth rate. Starting
from an initial condition of the TRN that corresponds to the
attractor of one of these 81 media, say X, we computed the time
course of the TRN configuration in another buffered medium Y, until
it reached the attractor corresponding to Y. For each of the TRN
configurations in the trajectory we computed the growth rate using
constrained FBA. This effectively tracks how the constrained growth
rate of the cell changes with time after its environment changes
suddenly from X to Y. The result is shown in Fig. 6 for the cases
where the carbon source in X is glutamate and in Y is glutamine,
lactate, fucose or acetate. In the attractor of X the growth rate is
low for the medium Y. The TRN configuration changes with time so as
to typically increase the growth rate.

We found that for the above 81 minimal media, the growth rate in the
attractor configuration of the medium was greater than the average
growth rate in the other 80 attractors by a factor of 3.5 (averaged
over the 81 media). Moreover the average time to move to the
attractor from such initial configurations was only 2.6 time steps.
In other words regulatory dynamics enables the cell to adapt to its
environment to increase its metabolic efficiency very substantially,
fairly quickly.

We also calculated the growth rate for each of the 15427 minimal
media in their respective attractor configurations as a ratio of the
maximal growth rate possible in those media (the latter computed for
each medium using FBA on the full metabolic network without imposing
any regulatory constraints). The average value of this ratio was
found to be as high as 0.815 and was less than 0.5 for only 7 \% of
the media (for the histogram of these ratios see Fig. 7). This shows
that the regulatory dynamics results in a close-to-optimal metabolic
functioning under a large set of conditions. This observation also
lends support to the usefulness of FBA in probing metabolic
organization.

%----------------------------------------------------------------------------------------------

\begin{figure}
\centering
\includegraphics[width=7cm]{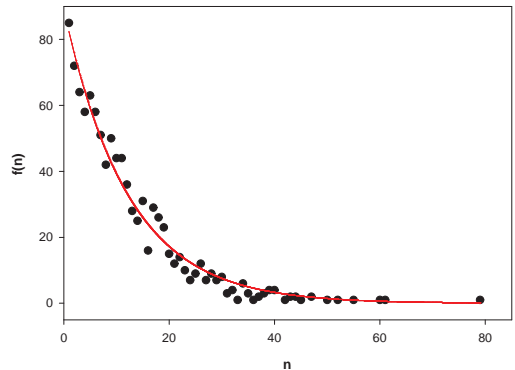}
\caption{Frequency distribution of the number of random knockouts
needed to make a cell unviable for growth for all 81 minimal media.
The red curve is the best fit to an exponential distribution.}
\end{figure}

\subsection*{Robustness of the network to gene knockouts}

In order to test the robustness of network functionality to
successive gene knockouts, we considered the progressive decline of
metabolic performance for an ensemble of 1000 `random knockout
trajectories'. Each trajectory was constructed as follows: One out
of 583 genes was chosen at random and knocked out, i.e., its $g_i$
was set to be identically 0. The constrained FBA growth rate was
determined for the attractors of the resultant dynamical system of
582 genes for each of the 81 minimal media discussed above. This was
repeated after knocking out another gene chosen at random from the
remaining 582 genes, and so on until the attractors for all the 81
media became dysfunctional (i.e., gave a zero growth rate). The
number of knockout steps, $n$, needed for the network to become
metabolically dysfunctional for all the 81 media was determined for
each of the 1000 random knockout trajectories constructed in this
way. Figure 8 shows the number or frequency $f(n)$ of trajectories
with a given value of $n$. The curve fits the exponential
distribution $f(n) \sim \exp(-n/n_0)$ with $n_0 = 12.1$. Thus the
chances of survival decrease exponentially with the number of
knockouts.

%----------------------------------------------------------------------------------------------

\subsection*{Design features of the regulatory network: Origin of
homeostasis and flexibility}

The following structural characteristics of the TRN explain several
of the dynamical features described above: The TRN 1A is an acyclic
directed graph with maximal depth 4. The largest connected component
is displayed as a hierarchy in Fig. 9, in which all links are
pointing downwards. At the bottom of the hierarchy are 479 metabolic
genes in the full system (409 in the largest connected component)
coding for enzymes that have no outgoing links. Thus these nodes do
not influence the dynamics of any other gene. We refer to these as
the `leaves' of the acyclic graph. At the top of the hierarchy are
nodes with no incoming links, or `root nodes'. The depth of a node
in the acyclic graph is the length of the longest path to it from a
root node. Root nodes correspond to external metabolites and other
variables that have fixed values in the system 1A such as certain
conditions, fluxes, etc. Since we consider only buffered media the
${\bf m}$ variables, by virtue of their root location, act as
control variables of the dynamical system. The genes coding for TFs
are at intermediate levels in the graph. These observations
immediately explain why (a) there are only fixed point attractors of
this system, (b) their basin of attraction is the entire
configuration space, (c) it takes at most 4 time steps to reach the
attractors from any initial configuration, and (d) the attractor
configuration depends upon the medium. For, the ${\bf m}$ vector
determines the configuration of the root level. This fixes the
configurations of all nodes at the next level (depth 1) at the next
time instant ($t=1$) and subsequent times irrespective of their
values at $t=0$, because the input variables to the boolean
functions controlling them are fixed. This fixes the configurations
of all nodes of depth 2 at $t=2$ irrespective of their
configurations at $t=1$, and so on, until at $t=4$, the
configuration of the maximum depth leaves are fixed irrespective of
the configuration they held earlier. A change in the medium or
external environment is a change in the configuration of root nodes;
this also percolates down in a maximum of 4 steps resulting in a new
fixed point. The acyclicity of the {\it E. coli} TRN was  noted by
\cite{SMMA2002}. Its maximum depth being 5 (including parts of the
network that regulate systems other than metabolism) was remarked
upon by \cite{MBZ2004}. That root control of this acyclic graph is
in the hands of environmental signals has been observed by
\cite{BBO2005}. However, to our knowledge the present work is the
first one that brings these facts together to study dynamics and
elaborate upon their consequences for homeostasis and flexibility of
the system.

%----------------------------------------------------------------------------------------------

\begin{figure*}
\centering
\includegraphics[trim=20 80 20 20,width=16cm]{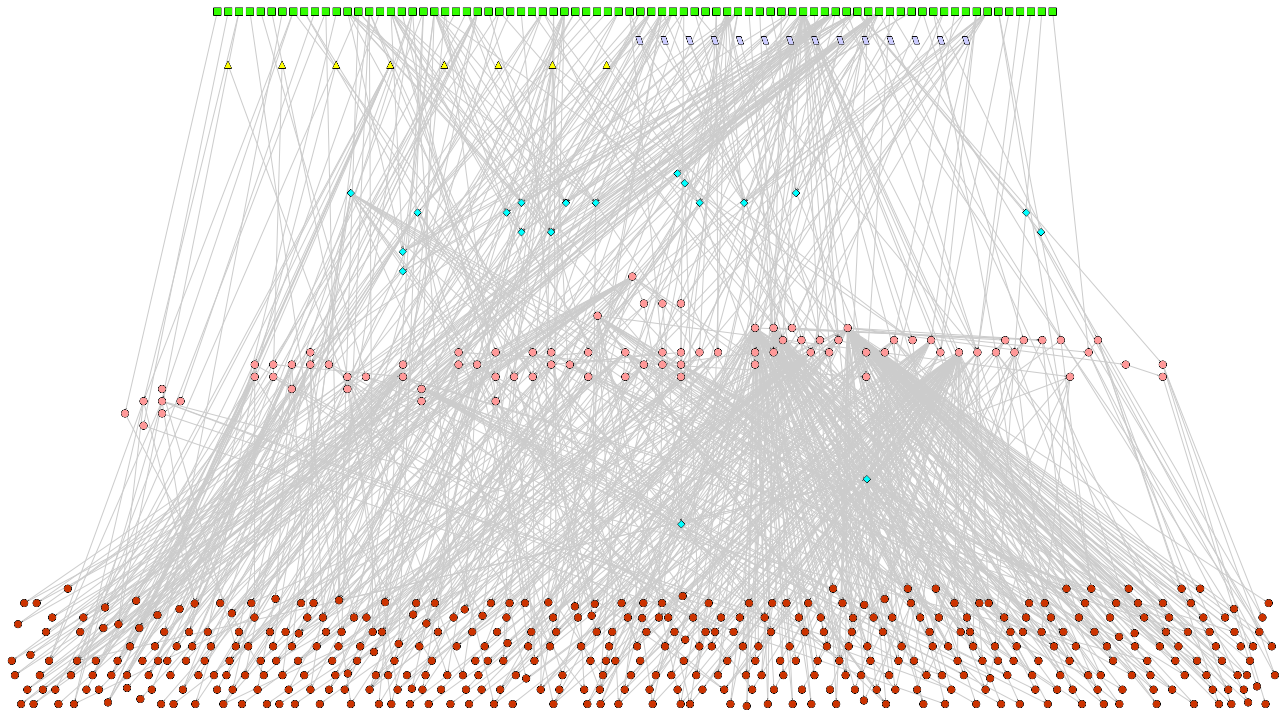}
\caption{Largest connected cluster of the TRN controlling metabolism
in {\it E. coli}. The colour coding of all nodes is as in Fig. 1. }
\end{figure*}

\subsubsection*{Disconnected structure of the reduced dynamical
system: modularity, flexibility and evolvability}

Since leaf nodes do not affect the dynamics of upstream nodes, it is
worthwhile to ask about the dynamics of the `reduced dynamical
system' which is obtained from the full system by removing the
leaves. When leaf nodes in the system are removed along with all
their links, one is left with Fig. 10. This is a surprisingly
disconnected graph; the large connected component has broken up into
38 disconnected components. It has several small components
containing upto only 4 nodes at depth $\geq 1$ and one component
with 27 nodes at depth $\geq 1$. The latter component is regulated
by oxygen, some inorganic sources of nitrogen, and certain amino
acids and sugars. Other components are typically regulated by single
metabolites or groups of biochemically related metabolites.  This
procedure reduces the number of outgoing links from global
regulators drastically. For example the gene b3357 coding for Crp is
left with only 3 outgoing links instead of 105.

Two components of a dynamical system that are disconnected from each
other are dynamically independent: the dynamics of each can be
analysed independently of the other. The dynamics of the full
system, in particular its attractors and basins of attraction, can
be reconstructed from those of its disconnected components. Such a
disconnected or `product' structure of a dynamical system greatly
simplifies its mathematical analysis. Modularity of biological
systems refers to the existence of subsystems that are relatively
independent of each other \cite{HHLM1999}. Each connected component
of Fig. 10 can therefore be regarded as a core of a module, and
modularity of the present genetic regulatory system is then nothing
but the property that it is composed of disconnected components at
this level of description.

Restoring the leaves and their links in Fig. 10 will take us back to
Fig. 1 which contains the large connected component shown in Fig. 9.
This means that leaf nodes typically receive links from more than
one module core. The structure is like a banyan tree which has
multiple trunks emanating from independent roots and in which leaves
receive sustenance from more than one root. In this picture, there
is no direct crosstalk between the module cores but they can affect
common leaves. This enables many leaf nodes to be influenced by
several environmental conditions. This `multitasking' adds to the
complexity of cellular response to different environments and
possibly contributes to greater metabolic efficiency. When a minimal
medium is changed by replacing its carbon source by another that
belongs to a different module, the genetic network needs to respond
by activating genes coding for enzymes that catalyze metabolic
reactions needed to break down the new source and process its
moieties. The connections of the leaf nodes to the modules above
them must be such that that is achieved, given our finding that the
constrained FBA growth rate increases as the new attractor is
reached.

\begin{figure*}
\centering
\includegraphics[trim=0 200 150 0,width=18cm]{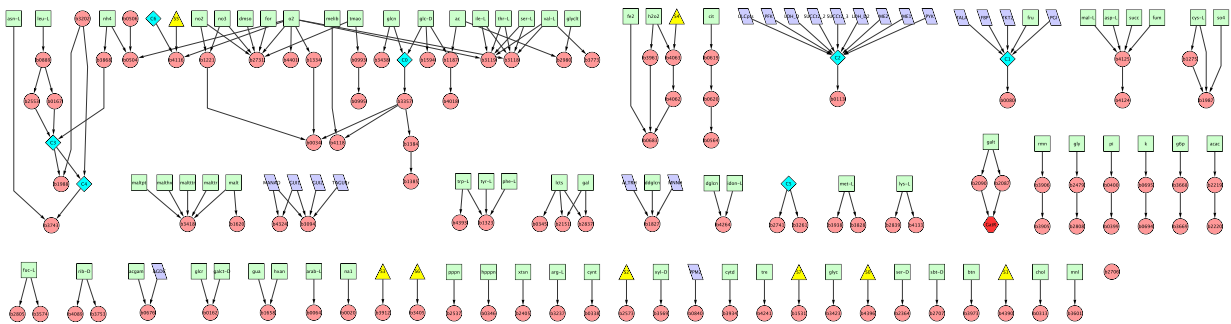}
\caption{Picture of the regulatory network obtained when all leaf
nodes in the network of Fig. 1 are removed along with all their
links. The colour coding of all nodes is as in Fig. 1. The red
hexagon denotes the lone TF in the network that is coded for by two
genes. The nomenclature for conditions C1 to C7 and S1 to S8 is
given in Table S3 in Additional File 1. The electronic version of
this figure can be zoomed in to read node names.}
\end{figure*}

The location and dynamical autonomy of the modules could also
contribute to evolvability. A new module added to Fig. 10 would not
affect existing ones; thus the organism can explore new niches
characterized by new food sources without jeopardizing existing
capabilities. This may be a particular case of the more general
observation \cite{AF2000,KG2005} that the architectural features of
organisms responsible for their flexibility to environmental
conditions also contribute to their evolvability.

The graph of the dynamical system 1B is not completely acyclic.
Effectively some of the genes that are leaves in 1A now get outgoing
links that feed back to genes coding for transcription factors. This
results in the cycles we have seen as attractors. Our analysis of
this dynamical system, not discussed here, reveals that removing the
leaves of this system exposes a modular structure in terms of which
the attractors can be understood.

%----------------------------------------------------------------------------------------------

\subsubsection*{Almost all input functions are canalyzing in the {\it
E. coli} TRN}

It has been shown by Kauffman and his colleagues that the stability
in the genetic regulatory networks to perturbations can arise due to
the canalyzing property of boolean functions
\cite{KPST2003,KPST2004}. A canalyzing boolean function has at least
one input such that at least one of the two values of this input
determines the output of the function \cite{OriginsOfOrder}. For a
given number of inputs, $K$, the fraction of boolean functions that
are canalyzing decreases as $K$ increases. All boolean rules
compiled for eukaryotes from the available literature have been
found to be canalyzing functions \cite{HSWK2002}. For the present
{\it E. coli} TRN the frequency distribution of the number of genes
with $K$ regulatory inputs is given in Table S4 in Additional File
1. We found that boolean functions for $579$ of the $583$ genes in
the {\it E. coli} TRN possess the canalyzing property. Only $4$
genes had input functions that were not canalyzing.

\subsubsection*{The dynamical system achieves flexibility even though it is
far from the edge of chaos}

One might expect that a dynamical system whose attractors have large
frozen cores and very small `twinkling islands' is rather rigid and
therefore unlikely to be adaptable to the external environment and
also unlikely to be evolvable. This expectation has given rise to
the conjecture (see \cite{OriginsOfOrder}) that genetic regulatory
systems ought to be close to the `edge of chaos', the boundary that
separates the ordered phase from the chaotic phase in the space of
dynamical systems. However, as discussed above in the section on
homeostasis, the present dynamical system is deep in the ordered
phase, since it always falls into the same attractor that is a fixed
point or has isolated low period cycles for all initial conditions
in a few time steps (all or most genes get frozen). In other words
it is far from the edge of chaos. We have seen that this is an
inevitable consequence of the hierarchical, largely acyclic
architecture of the network (see the section on design features). At
the same time, we have seen that the system is also highly
responsive to the environment. How have these two properties managed
to co-exist? The answer lies in the observation that root nodes of
the hierarchy are largely the environmental variables -- the
external metabolites in the present case. The attractor
configuration is thus a function of the external environment,
specified by the variable ${\bf m}$. While for any fixed ${\bf m}$
there is a global attractor in which most or all genes have frozen
configurations, when ${\bf m}$ changes the genes `unfreeze' and move
to a new attractor configuration. The modular organization of the
network with a lot of crosstalk between modules at the leaf level
(enzyme coding genes) ensures that the melting and refreezing is
quite substantial. The same architecture that produces this
flexibility of response to the external environment can also enhance
evolvability.

The present architecture as an alternative to the edge of chaos
hypothesis for simultaneously producing homeostasis and flexibility
has not been noticed earlier because the earlier literature has
primarily focussed on the abstract genetic network itself without
much reference to the environmental control variables that abound in
the real systems. Here, since we are investigating the database
iMC1010v1 which brings together, within the same network, genes as
well as nodes describing external environmental signals, this
possibility has become evident.
%----------------------------------------------------------------------------------------------

\section*{Discussion and Conclusions}

The overall organizational picture of the system that emerges from
our study is the following: The network's largely acyclic structure
means that there must exist nodes that have no incoming links from
within the network (root nodes, sitting at the top of the
hierarchy). If the configuration of the root nodes is held fixed,
the network dynamics necessarily flows to a fixed point attractor
whose configuration is insensitive to the initial configuration or
perturbation of the rest of the nodes. It turns out that the root
nodes are typically the external metabolites and other environmental
conditions, while the rest of the network consists of genes. Thus
the system simultaneously possesses the two properties that
attractor configurations are (a) insensitive to initial gene
configurations or their perturbations (homeostasis) and (b)
sensitive to external environmental conditions (flexibility of
response).

Acyclicity also means that there are leaf nodes (with no outgoing
links to the network) at the bottom of the hierarchy. Deleting the
leaf nodes along with their links reveals the remainder of the
network to be consisting of a large number of disconnected
components; see Fig. 10. By construction each of these components is
a subsystem whose dynamics is independent of the rest of the system.
An aspect of modularity of a system is the dynamical autonomy of
certain subsystems; this property is thus clearly visible. Most
modules are controlled at the root by a set of biochemically related
metabolites or a single metabolite.

All our results, being derived from the database iMC1010v1, have
some limitations that stem from the database itself. First, the
database covers the regulation of only about half of the metabolic
genes in {\it E. coli}. Even among these genes the present set of
connections could have false positives as well as negatives,
especially the latter. Additional nodes and connections would modify
the dynamics reported here. However, new nodes and connections
corresponding to genes coding for enzymes are unlikely to affect our
qualitative conclusions about the nature of attractors
significantly. The reason is that most such genes are likely to be
leaves of the network like the nodes at the bottom of Fig. 9, in
which case they would not affect the dynamics of other nodes.
However the inclusion of such genes as well as additional
connections of existing genes in the network would add to the
constraints on FBA; it would be interesting to see the extent to
which regulatory dynamics enhances metabolic efficiency in different
environmental conditions.

The inclusion of more TF genes and modified connections among
existing genes would affect the dynamics. In particular feedback
loops could bring in longer cycles as attractors. Several genes are
known to have autoregulatory self-loops \cite{RegulonDB} that are
not included in the present database. These could produce 2-cycles
at the individual nodes even at constant input. Present work seems
to indicate that apart from self-loops, TRNs are largely acyclic
\cite{SMMA2002,MBZ2004,BBO2005} and have a small depth (about 5).
Furthermore the kind of modularity described here for the TRN
regulating metabolism seems to exist for other parts of the {\it E.
coli} TRN. This together with the evidence of preponderance of
canalyzing functions suggests that cyclic attractors where they do
exist are likely to be of low period and localized. Cyclicity is
needed for explicitly temporal phenomena like the cell cycle or
circadian rhythms. It is possible that metabolism being a
functionality that needs to be active whenever food is available is
largely regulated without cycles at the genetic level, with
feedbacks typically entering at the level of metabolites regulating
enzymes to ensure efficient functioning on a faster time scale.
Nevertheless it would be important to explore these questions with
an enlarged database.

We end with a comment relating this to earlier works and a
speculation. Kauffman \cite{K1969a,OriginsOfOrder} has found
biologically motivated random boolean networks to possess multiple
attractors that he has interpreted as different cell types of a
multicellular organism. In the present work, we have studied the
genetic network regulating metabolism in a prokaryote. Perhaps not
surprisingly, we get a much simpler picture of the network
exhibiting a much higher degree of order in the dynamics than the
systems Kauffman investigated. While we also find that the system
can go into different attractors (see the discussion above on
flexibility), yet, unlike Kauffman, for whom different attractors
were realized via different initial conditions of the genes, in the
present case the different attractors are realized when the control
variables (metabolites in the external environment) have different
configurations. When the control variables are held fixed we find no
(or very little) multiplicity of attractors irrespective of the
initial condition of the genes (see the discussion on homeostasis).
This architecture and dynamics is probably quite suitable for
prokaryotic lifestyles and evolution. The question remains open
whether for eukaryotes and especially multicellular ones, Kauffman's
hypothesis that associates different cell types with different
attractors of the regulatory dynamics is valid. While that
hypothesis remains an enticing possibility, it is worth noting that
the present simple architecture would have its uses in the
eukaryotic case as well. Environmental control of cellular
attractors (via the architecture discussed above) can itself cause a
cell to differentiate into another type, the environment being
determined in the multicellular case by the state of other cells in
the organism. The modular structure discussed above would even
permit a cell to hop through several attractors in the course of
development of the organism as the environmental cues to this cell
change. Such an architecture could thus contribute to developmental
flexibility and, potentially, evolvability of eukaryotes as well.
The multiplicity of internal attractor basins pointed out by
Kauffman would be an asset in keeping the cell in its new attractor
after a transient environmental cue has caused it to shift from one
basin to another. It would be interesting to investigate these
questions when a database similar to iMC1010v1 becomes available for
a multicellular organism.

%----------------------------------------------------------------------------------------------

\section*{Methods}

%----------------------------------------------------------------------------------------------

\subsection*{Construction of the boolean dynamical system describing
the genetic regulation of {\it E. coli} metabolism}

We have represented the {\it E. coli} TRN regulating its metabolism
as a boolean dynamical system given by the equation (1) where
$g_i(t)$ represents the configuration of gene $i$ (with values 0 or
1 representing the gene being off or on, respectively) at time $t$,
and the vector ${\bf m} = (m_1,\ldots,m_{96})$ describes the
buffered external environment ($m_i$ being 0 or 1 if metabolite $i$
is absent or present, respectively, in the external environment).
This dynamical system was constructed from the integrated regulatory
and metabolic network iMC1010v1 for {\it E. coli} \cite{CKRHP2004}.
This database was downloaded from the website
\url|http://gcrg.ucsd.edu/|. The regulatory interactions and the
boolean rules incorporated in this reconstructed network are based
on various literature sources. The TRN accounts for $583$ genes of
which $479$ are coding for enzymes catalyzing metabolic reactions
and $104$ are coding for TFs. The 583 genes, 103 TFs, 96 external
metabolites, 19 conditions, and 21 internal fluxes of metabolic
reactions are respectively denoted by the vectors ${\bf g,t,m,c,v}$,
all of which can, in principle, depend upon time $t$. E.g., $t_i(t)$
($i=1,2,\ldots,103$), the $i^{th}$ component of ${\bf t}(t)$, equals
unity if the TF $i$ is present in the cell at time $t$ and zero if
not. $c_i(t)$ ($i=1,2,\ldots,19$), the $i^{th}$ component of ${\bf
c}(t)$, equals unity if the $i^{th}$ condition holds at time $t$ and
zero if not. $v_i(t)$ ($i=1,2,\ldots,21$), the $i^{th}$ component of
${\bf v}(t)$, equals unity if the $i^{th}$ metabolic reaction in the
above mentioned set of internal metabolic reactions is happening
inside the cell at time $t$ (with a flux greater than a specified
value) and zero if not. The additional 9 stimuli (e.g. stress, etc.)
are assumed to be absent. Thus the overall system contains
583+103+96+19+21=823 boolean variables. Its dynamics is organized as
follows: The presence or absence of the transcription factors,
external metabolites, and the status of the internal fluxes and
other conditions at time $t$ determines the on-off state of the 583
genes at $t$:
\begin{equation}
g_i(t) = F_i({\bf t}(t),{\bf m}(t),{\bf c}(t),{\bf v}(t)), \quad
\quad i = 1,2,\ldots,583.
\end{equation}
The database iMC1010v1 gives the form of the functions $F_i$ in
terms of AND, OR and NOT operations on the boolean arguments. The
103 transcription factors are coded for by a subset of 104 genes
(two genes together code for one TF and the remaining 102 genes code
for one TF each). The on-off state of these genes at the previous
time step $t-1$ determines whether the TFs they code for are present
at $t$ (a single time step therefore corresponds to the average time
for transcription and translation). Thus
\begin{equation}
t_i(t) = T_i({\bf g}(t-1)), \quad \quad i = 1,2,\ldots,103,
\end{equation}
where the function $T_i({\bf g}) = g_i$ for 102 transcription
factors that are coded for by single genes; for the TF coded for by
2 genes $T_i({\bf g}) = g_{i_1} \; AND \; g_{i_2}$. Substituting
this in the previous equation gives
\begin{equation}
g_i(t) = F_i({\bf T}({\bf g}(t-1)),{\bf m}(t),{\bf c}(t),{\bf
v}(t)).
\end{equation}
This equation provides the dynamical rule for updating the gene
configurations from one instant to the next, provided the status of
the variables ${\bf m,c,v}$ is known.

%----------------------------------------------------------------------------------------------

\subsection*{Treatment of external metabolites ${\bf m}$}

In this work we considered only buffered media in which the external
environment was assumed constant. Thus ${\bf m}(t)= {\bf m}$,
independent of $t$. For each medium considered, the components of
{\bf m} corresponding to the metabolites present in the external
environment were set to unity and the remaining components were set
to zero. {\it E. coli} is known to be capable of transporting 143
metabolites into the cell, including 131 organic and 12 inorganic
molecules \cite{RVSP2003} of which 96 (86 organic and 10 inorganic)
are included in the regulatory part of the database iMC1010v1. We
considered the following classes of minimal media in this work:\\
(a) 93 minimal media (61 aerobic and 32 anaerobic): These are
characterized by a single organic source of carbon (listed in
Supplementary Table S1), and the ions of ammonium, sulphate,
phosphate, hydrogen, iron, potassium and sodium. The components of
{\bf m} corresponding to these metabolites were set to unity and
others were set to zero in a given minimal medium. Oxygen was set to
unity in the aerobic media and to zero in anaerobic media. In
principle 86 organic carbon sources would yield 172 media (aerobic
plus anaerobic). Out of these we restricted ourselves to that subset
of media for which the {\it E. coli} metabolic network supports
growth of the cell as determined by Flux Balance Analysis (FBA);
i.e., media for which the optimal growth rate calculated by FBA
without imposing regulatory constraints is nonzero (see below). This
condition yielded the list of 93 media listed in Supplementary Table
S1. Most of the work reported in this paper was performed with this
set of
minimal media.\\
(b) For part of our work we also considered a much larger library of
minimal media, described by \cite{BHRP2005}, in which all possible
combinations of single sources of carbon, nitrogen, sulphur,
phosphorus, etc., from among the 143 metabolites ingested by {\it E.
coli} are considered. Following the method described in the
supplementary material of \cite{BHRP2005}\ gave us a library of
109732 minimal media.

%----------------------------------------------------------------------------------------------

\subsection*{Treatment of the conditions ${\bf c}$ and internal
fluxes ${\bf v}$}

\noindent {\it The ${\bf  c}$ variables:} Of the 19 boolean
variables $c_i(t)$, 15 depend only on the configuration of a subset
of TFs and external metabolites at time $t$, i.e., $c_i(t) =
C_i({\bf t}(t),{\bf m}(t))$, $i=1,2,\ldots,15$, where the $C_i$ are
specified boolean functions in the database. These functions can be
substituted in Eq. 2. This eliminates these 15 variables $c_i$ from
the dynamical system at the expense of a more complicated effective
dependence of $g_i(t)$ on ${\bf t}(t)$ and ${\bf m}$. Of the
remaining 4 conditions, one, representing growth of the cell, is set
to unity (since we primarily consider only those conditions in which
the cell has a nonzero growth rate). Another condition represents
the pH of the external environment, which we take to be between 5.5
and 7 (weakly acidic, as, for example, in the human gut). The pH
condition affects only 3 genes in the database. For two of them the
operative regulatory clause is `pH $<$ 4'; we take the boolean
variable $c_i$ corresponding to pH to be zero (false) for these two
genes. For the third gene the clause is `pH $<$ 7'; for this gene we
take this variable to be unity (true). Two other conditions,
designated as `surplus FDP' and `surplus PYR' in the database,
correspond to whether `surplus' amounts of fructose 1,6-bisphosphate
and pyruvate are being produced in the cell. These conditions depend
upon the values of some of the internal fluxes $v_i$ and the
presence of an external metabolite, fructose, through specified
boolean functions. The latter variable is treated as unity if the
minimal medium includes fructose and zero otherwise, as discussed
above. The treatment of the internal fluxes is discussed below.\\

\noindent {\it The ${\bf v}$ variables:} The 21 components of the
vector ${\bf v}$ represent fluxes of 21 metabolic reactions. As
mentioned by \cite{CKRHP2004}, these are surrogate for other
conditions inside the cell, e.g., concentrations of metabolites
produced by those reactions, which can affect gene
regulation. We have treated these variables in two distinct ways. \\
(A) In the first approach we identified whether the particular
metabolic reaction was a `blocked reaction' or not
\cite{SS1991,BNSM2004,SSGKRJ2006}. A reaction is said to be blocked
in a particular environmental condition (specified by a buffered
medium) if under that medium no steady-state flux is possible
through it \cite{BNSM2004}. This can be determined using metabolic
flux analysis methods from a knowledge of the metabolic network. For
each medium (specified by the vector ${\bf m}$) we chose the fixed
value zero for a particular flux variable $v_i(t)$ if it was found
to be blocked for that condition, and unity otherwise. Thus in this
approach the $v_i$ were not dynamical variables, but rather
fixed parameters (albeit fixed with an eye on self-consistency). \\
(B) In the second approach, we allowed the $v_i$ to be dynamical,
but made a simplifying assumption about their dynamics. In the cell,
the flux values of individual reactions are determined by the
concentrations of participating metabolites and the catalyzing
enzymes, the latter being controlled by the activity of their
respective genes. In a discrete-time approximation, an enzyme is
present at time $t$ if the genes coding for it are active at $t-1$.
Thus we set $v_i(t) = 1$ if the genes coding for the enzyme of that
metabolic reaction were active at $t-1$, and $v_i(t) = 0$ otherwise.
This could be done for a subset of 10 out of 21 reactions, since the
genes of their enzymes were part of the 583 genes in the database.
Genes coding for the enzymes of the remaining 11 reactions were not
part of the database and hence the corresponding $v_i$ could not be
made dynamical variables in this fashion. These latter $v_i$ were
fixed as in part (A) for each medium. The approach (B) introduces
feedbacks in the genetic regulatory network.

Our above treatment defines the substitutions to be made in Eq. 2
for the variables ${\bf c}(t),{\bf v}(t)$. Each component of {\bf c}
in Eq. 2 is either a specified boolean function of ${\bf t}(t)$,
{\bf m}, and ${\bf v}(t)$, or is a suitably chosen boolean constant.
Each component of ${\bf v}(t)$ is, in turn, either a specified
boolean function of ${\bf g}(t-1)$, or is a suitably chosen boolean
constant. These substitutions together with Eq. 3 make the right
hand side of Eq. 2 a function of only ${\bf g}(t-1)$ and ${\bf m}$,
i.e., Eq. 2 reduces to $g_i(t) = G_i({\bf g}(t-1), {\bf m})$, which
is the same as Eq. 1. The functions $G_i$ define the final dynamical
system, and include information coming from the functions $F_i$, as
well as the dependence of ${\bf t}$, ${\bf c}$ and ${\bf v}$ on
${\bf g}$ and ${\bf m}$. Note that the choices (A) and (B) for the
${\bf v}$ variables yield different dynamical systems for Eq. 1
which we denote as 1A and 1B respectively; in 1B 6 out of 583 genes
have additional links from other genes in the set compared to 1A.
Programs implementing these two dynamical systems are available from
the authors.

%----------------------------------------------------------------------------------------------

\subsection*{Computation of Growth rate of {\it E. coli} for a
given environmental condition}

\noindent Flux Balance Analysis (FBA) is a computational technique
that determines the maximal steady state growth rate of a cell that
its metabolic network can support in any given buffered medium
\cite{VP1994,EP2000,SVC2002}. The database iMC1010v1
\cite{CKRHP2004} includes the {\it E. coli} metabolic network
database iJR904 \cite{RVSP2003} to which FBA can be applied. In
this work we use FBA in two ways:\\

\noindent {\it Pure (unconstrained) FBA.} This uses the full
metabolic network iJR904 (without any constraints from regulation)
to calculate the maximal growth rate of the cell under various
media. A zero value of the maximal growth rate for a particular
medium means that the metabolic network does not contain pathways to
convert the substances present in the medium into `biomass
metabolites' needed for cell growth.\\

\noindent {\it FBA with regulatory constraints.} Of the 583 genes in
the database iMC1010v1 479 genes code for enzymes of the metabolic
reactions in the database iJR904. In any given configuration of the
genetic network a subset of these genes is off and the remaining are
on. Thus one can run FBA wherein those reactions of the metabolic
network are switched off whose enzymes are not being produced (i.e.,
whose corresponding genes are off). We will refer to this as
`constrained FBA'. In this way one can track the optimal growth rate
as a function of time as the configuration of the genes changes
according to the dynamics of the genetic regulatory network, as
discussed in \cite{CSP2001,CKRHP2004}. The growth rate obtained from
constrained FBA for any configuration of the genes is, by
definition, less than or equal to that obtained from pure FBA (for
the same medium).

%-------------------------------------------------------------------
\section{Acknowledgement}

We thank Bernhard O. Palsson and Jennifer L. Reed for providing the
GPR association of the {\it E. coli} metabolic network. We would
like to thank Devapriya Choudhury, Varun Giri, Shobhit Mahajan, N.
Raghuram, Anirvan Sengupta and Shalini Singh for discussions. We
also thank the Bioinformatics Centre, University of Pune for
infrastructural support. Graph visualization has been done using
Cytoscape. A.S. would like to acknowledge a Senior Research
Fellowship from Council of Scientific and Industrial Research,
India. The work is supported by a grant from Department of
Biotechnology, Government of India.

%-------------------------------------------------------------------

%-------------------------------------------------------------------

%-------------------------------------------------------------------
\end{document}